%% file: ISIT15_FD.tex
\documentclass[conference]{IEEEtran}
\usepackage{amsmath,amssymb,mathrsfs,cite}
\usepackage{epsfig,epsf,subfigure,graphicx,graphics}
\usepackage{url}

\input{commands.tex}

\graphicspath{{./figs/}}

\IEEEoverridecommandlockouts

\allowdisplaybreaks

\title{Opportunistic Scheduling for Full-Duplex Uplink-Downlink Networks}
\author{
\authorblockN{Can Karakus}
\authorblockA{
UCLA, Los Angeles, USA\\
\textsf{karakus@ucla.edu}}
\and
\authorblockN{Suhas Diggavi}
\authorblockA{
UCLA, Los Angeles, USA\\
\textsf{suhasdiggavi@ucla.edu}}
\thanks{This work was supported in part by NSF grant \#1314937 and a gift by Intel Corp.}
}

\begin{document}
\maketitle
\begin{abstract}
We study opportunistic scheduling and the sum capacity of
cellular networks with a full-duplex multi-antenna base station and a
large number of single-antenna half-duplex users. Simultaneous uplink
and downlink over the same band results in uplink-to-downlink
interference, degrading performance. We present a simple opportunistic
joint uplink-downlink scheduling algorithm that exploits multiuser
diversity and treats interference as noise. We show that in
homogeneous networks, our algorithm achieves the same sum capacity as
what would have been achieved if there was no uplink-to-downlink
interference, asymptotically in the number of users. The algorithm
does not require interference CSI at the base station or uplink
users. It is also shown that for a simple class of heterogeneous
networks without sufficient channel diversity, it is not possible to
achieve the corresponding interference-free system capacity. We
discuss the potential for using device-to-device side-channels to
overcome this limitation in heterogeneous networks.
\end{abstract}

\section{Introduction}
\input{Intro.tex}

\section{Model and Notation}\label{sec:model}
\input{Model.tex}

\section{Sum Capacity in Homogeneous Networks}\label{sec:homo}
\input{Homo.tex}

\section{Sum Capacity in Heterogeneous Networks}\label{sec:hetero}
\input{Hetero.tex}

\bibliographystyle{ieeetr}
\bibliography{Ref}

\appendices
\section{Proof of Theorem~\ref{th:homo}}\label{ap:proof_homo}
\input{AppendixHomoPf.tex}

\section{Proof of Theorem~\ref{th:antennascaling}} \label{ap:antennascaling}
\input{AppendixAntennaScaling.tex}

\section{Proof of \eqref{eq:upperbound}}\label{ap:upperbound}
\input{AppendixUpperBound.tex}

\section{Proof of Corollary~\ref{cor:gap} }\label{ap:gap}
\input{AppendixHeteroGap.tex}

\section{Auxiliary Lemmas}\label{ap:lemmas}
\input{AppendixLemmas.tex}

\end{document}

%% file: commands.tex
\newtheorem{lemma}{Lemma}[section]
\newtheorem{theorem}{Theorem}[section]

\newtheorem{corollary}{Corollary}[section]

\newtheorem{remark}{Remark}[section]

\newcommand{\Prob}[1]{\mathbb{P} \left( #1 \right)}
\newcommand{\MAC}{\text{MAC-M}}
\newcommand{\BC}{\text{BC}}
\newcommand{\SNR}{\mathsf{SNR}}

\newcommand{\aaaa}{\mathrm{(a)}}
\newcommand{\bbbb}{\mathrm{(b)}}
\newcommand{\cccc}{\mathrm{(c)}}
\newcommand{\dddd}{\mathrm{(d)}}
\newcommand{\eeee}{\mathrm{(e)}}
\newcommand{\ffff}{\mathrm{(f)}}
\newcommand{\gggg}{\mathrm{(g)}}
\newcommand{\hhhh}{\mathrm{(h)}}
\newcommand{\iiii}{\mathrm{(i)}}

\newcommand{\lp}{\left(}
\newcommand{\rp}{\right)}
\newcommand{\lb}{\left[}
\newcommand{\rb}{\right]}
\newcommand{\lbp}{\left\{}
\newcommand{\rbp}{\right\}}

\newcommand{\wtild}{\widetilde}

%% file: Intro.tex
Full-duplex wireless communication is becoming closer to reality, in
light of recent experimental results demonstrating its
feasibility \cite{DuarteSabharwal_10, ChoiJain_10}. Especially the
development of massive MIMO can create opportunities for full-duplex
communication, since all implementations of full-duplex use multiple
antennas. The first application of full-duplex in a practical system
is expected to be in base stations instead of mobile devices, due to
relative flexibility in design. Since mobile devices remain
half-duplex, the uplink-downlink nature of a cellular system is
retained, even when the base station is full-duplex. By serving uplink
and downlink simultaneously over the same band, a full-duplex cellular
system might have the potential to double the spectral
efficiency. However, in order to realize this gain, one is immediately
faced with a challenge that is not present in half-duplex systems:
uplink-to-downlink interference.

The problem of uplink-to-downlink interference management in
full-duplex systems has been considered in \cite{SahaiDiggavi_13}
and \cite{BaiSabharwal_13} with several interference management
strategies proposed, based on interference alignment or message
splitting. However, in a large network, such sophisticated
interference management strategies can be impractical, may require
tight coordination between nodes and a large amount of
device-to-device CSI feedback. 

In this paper we explore schemes that require much less CSI overhead
by proposing a combination of opportunistic beamforming along with
treating uplink-to-downlink interference as noise. This enables us to
design an opportunistic joint uplink-downlink scheduling algorithm
that, in a homogeneous network with a large number of half-duplex
users and a multi-antenna full-duplex base station, asymptotically
achieves the sum of the capacities of the isolated uplink and downlink
systems, thus doubling the spectral efficiency. The main idea
underlying the result is to apply random transmit and receive
beamforming at the base station \cite{ViswanathTse_02}, and exploit
the multiuser diversity in the system to schedule the uplink and
downlink users that conflict the least with each other.

Many authors (including\cite{SharifHassibi_05, YooGoldsmith_06}, among
others) have studied the problem of MIMO downlink scheduling in
the many-user regime, and it has been demonstrated that the same scaling law as the
optimal dirty-paper coding sum rate can be achieved via beamforming
with scheduling. It was also shown that the gap between the sum rate
achievable with beamforming with scheduling and dirty-paper coding
goes to zero \cite{BayestehKhandani_08, WangLove_08}. There has also been works that explore how to exploit multiuser diversity in the presence of interference, under multi-cell downlink \cite{LiLiu_06}, and spectrum sharing cognitive radio \cite{BanChoi_09} scenarios. However, the
schemes developed in these works are either intended for an isolated downlink
system, or fail to provide any theoretical performance guarantees on the overall system throughput when translated into a full-duplex system, where the goal is to \emph{simultaneously} extract uplink and
downlink multiuser diversity gains while dealing with the
uplink-to-downlink interference. Based on the existing literature on
opportunistic scheduling, it is not clear whether
downlink sum rate optimality through scheduling is maintained in the
presence of uplink interference, especially when the uplink sum rate optimality is also sought.

We have two main contributions in this work. First, we show that the
asymptotic sum rate optimality in both uplink and downlink can be
maintained individually, even in the presence of uplink-to-downlink
interference. To achieve this, we develop a simple opportunistic
scheduling algorithm based on random beamforming. The algorithm does
not require the base station or the uplink users to have channel
information about the interference links. Moreover, very little CSI is
required at the base station due to random beamforming. We also show
that the spatial multiplexing gain offered by the multiple antennas is
retained in the full-duplex system when the number of antennas scale
logarithmically with the number of users, as was shown for
isolated downlink in \cite{SharifHassibi_05}.

This asymptotic decoupling result relies on there being sufficient
channel diversity in the network. Although a homogeneous network with
i.i.d. fading links provides sufficient diversity for this purpose,
such diversity may not be present in a real network. For instance,
there might be areas in a cell where users are densely clustered, and
some other areas that are mostly deserted, resulting in a lack of
sufficiently rich channel conditions. In a full-duplex system, in
addition to diversity in channels to and from the base station,
diversity in interference links is also required to realize the
multiuser diversity gains. Our second contribution is to show that for
a simple class of heterogeneous networks, it is not possible to
achieve such gains, by deriving an upper bound on the achievable sum
rate. In particular, the gap between the achievable sum rates of the
full-duplex system and the decoupled system grows linearly with the
number of antennas and logarithmically with downlink $\SNR$. Although
our heterogeneous network model is rather simple, it features the key
property of the lack of channel diversity. To address this limitation
in heterogeneous networks, we demonstrate through an example that
establishing device-to-device cooperation over orthogonal
side-channels can be effective.


%% file: Model.tex
\begin{figure}
\centering
\includegraphics{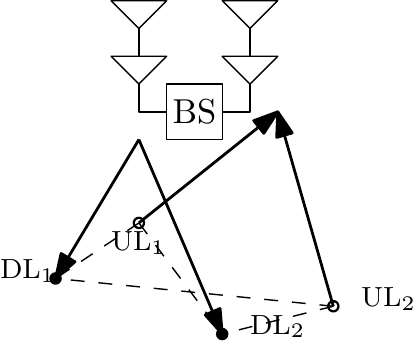}
\caption{A cellular system with a full-duplex base station with $M=2$ antennas and $n=2$ uplink and downlink half-duplex users. Uplink users are represented with white dots, downlink users are represented with black dots, and the interference links are represented with dashed lines.}
\label{fig:homo}
\end{figure}

We consider a cellular system with a single full-duplex base station, equipped with $M$ antennas for uplink and $M$ antennas for downlink communication (see Figure 1). We assume there are $n$ uplink, and $n$ downlink half-duplex users, each with a single antenna, requesting communication over the same band. We assume the base station is able to completely cancel self-interference, but the uplink transmission interferes with the received signal at the downlink users.

We first consider a homogeneous network, where all links in the network, including the interference links, are are generated i.i.d. from a $\mathcal{CN}(0,1)$ distribution; but once drawn, they remain fixed throughout the duration of transmission.

The uplink channel is described by the equation
\begin{align*}
\bar y = \bar H_n \bar x + \bar z,
\end{align*}
where $\bar y \in \mathbb{C}^{M\times 1}$ is the vector of channel outputs at the base station, $\bar x \in \mathbb{C}^{n\times 1}$ is the vector of channel inputs from $n$ uplink users, subject to a per-user block power constraint $\frac{1}{T}\sum_{t=1}^T \left| \bar x_k [t]\right|^2 \leq \bar P$ for a block length of $T$, for $k=1,\dots,n$,
$$\bar H_n=\lb\begin{array}{ccc}\bar h_1&\dots&\bar h_n\end{array}\rb \in \mathbb{C}^{M\times n}$$ is the matrix of channel gains, with each element generated i.i.d. according to $\mathcal{CN}(0,1)$, and $\bar z \sim \mathcal{CN}(\mathbf{0},\mathbf{I}_M)$ is the vector of complex Gaussian noise. Throughout the paper, we use the bar notation whenever a variable pertains to the uplink transmission, whereas we use plain letters for variables pertaining to downlink transmission, including the uplink-to-downlink interference link gain.

The downlink of the system is described by
\begin{align*}
y = H_n^* x + G_n \bar x + z,
\end{align*}
where $ y \in \mathbb{C}^{n\times 1}$ is the vector of channel output at the $n$ downlink users, $ x \in \mathbb{C}^{M\times 1}$ is the vector of channel inputs from $M$ antennas, subject to a total block power constraint $\frac{1}{T}\sum_{t=1}^T x^*[t] x[t] \leq P$, $H_n \in \mathbb{C}^{n\times M}$ is the matrix of channel gains and $G_n \in \mathbb{C}^{n\times n}$ is the matrix of interference link gains, with each element of the matrices generated i.i.d. according to $\mathcal{CN}(0,1)$, and $ z \sim \mathcal{CN}(\mathbf{0},\mathbf{I}_n)$ is the vector of complex Gaussian noise.

The set of all link gains in the network is denoted by $\mathcal{H}_n = \left( \bar H_n, H_n, G_n\right)$. Further, the rate of $i$th downlink (uplink) user is denoted by $R_i \left( \mathcal{H}_n\right)$ ($\bar R_i \left( \mathcal{H}_n\right)$), and the sum uplink and downlink rates are denoted by
\begin{align*}
\mathbf{\bar R}_n \left( \mathcal{H}_n\right) = \sum_{i=1}^n \bar R_i\left( \mathcal{H}_n\right),\;\;\;\;&\mathbf{R}_n \left( \mathcal{H}_n\right) = \sum_{i=1}^n R_i\left( \mathcal{H}_n\right).
\end{align*}
All logarithms throughout the paper are assumed to be in base $e$. We also define
\begin{align*}
\lb n\rb := \lbp k\in\mathbb{N}: 1\leq k \leq n \rbp.
\end{align*}

We impose the constraint that at most $M$ uplink users can simultaneously transmit to the base station, \emph{i.e.}, the vector $\bar x$ can only have $M$ non-zero elements per time slot\footnote{This constraint is placed to prevent total uplink power in the system from growing unboundedly.}.

%% file: Homo.tex
\subsection{Opportunistic Scheduling}

We consider an opportunistic scheduling algorithm that performs random beamforming \cite{ViswanathTse_02} independently for uplink and downlink, and schedules the users whose channels best fit to the current beamforming patterns, and least interfere with each other. In particular, the base station first constructs a random unitary matrix $\bar \Phi$ and multiplies this with the received uplink channel output
\begin{align*}
\bar \Phi^* \bar y &=\bar \Phi^* \bar H_n \bar x + \bar\Phi^* \bar z
\end{align*}
Note that since $\bar\Phi$ is unitary, $\bar\Phi^* \bar z$ is still distributed as $\mathcal{CN}(\mathbf{0},\mathbf{I}_M)$.
We consider the scheduling of $M$ uplink users for transmission at a given time. In particular, each element of the vector $\bar \Phi^* \bar y$ is assigned to a user, and the signal of that user is decoded from this component of the effective channel output, treating inter-stream interference as noise\footnote{Although successive cancellation decoding can also be used, this does not improve our main result, hence we treat interference as noise for simplicity.}. Note that this can be viewed as choosing an $M\times M$ submatrix of $\bar \Phi^* \bar H_n$. We use the following rule to choose the user $U_m \in \{ 1,\dots,n\}$ assigned to the $m$th stream:
\begin{align*}
\bar U_m = \arg\min_{k \in \bar S_m} \left| \bar \phi_m^* \bar h_k\right|^2 
\end{align*}
where
\begin{align*}
\bar S_m = \{ 1\leq k \leq n: \left| \bar \phi_m^* \bar h_k\right|^2 \leq \epsilon_n, \forall r \neq m \}
\end{align*}
for some $\epsilon_n$ such that $\epsilon_n\to 0$ as $n\to\infty$\footnote{Note that $\epsilon_n$ must be scaled down slow enough to ensure that $\left| \bar S_m\right|>0$ with high probability. The exact scaling of $\epsilon_n$ is left unspecified here, but in the proof of our main result, it will be seen that $\epsilon_n=O\lp \frac{1}{\log n}\rp$ is a good choice.}, where $\bar \phi_m$ is the $m$th column of $\bar \Phi$. Note that this scheduling algorithm first determines a set of candidate users for stream $m$, by eliminating all users whose interference to any other stream exceeds a certain threshold, and then picks the user whose channel has the largest projection along the $m$th beamforming vector in the candidate set. We denote the set of uplink users scheduled in this way as $\mathcal{\bar T} = \{ \bar U_m\}_{m=1}^M$.

Next, we consider the scheduling of downlink users, based on the uplink user selection. As in the uplink case, we begin by generating a random beamforming matrix $\Phi$, and precode the transmitted signal with it, so that the vector of received signals at the $n$ downlink users becomes
\begin{align*} 
y = H_n^* \Phi x + G_n \bar x + z,
\end{align*}
We use the following rule to choose the user $ U_m \in \{ 1,\dots,n\}$ assigned to the $m$th stream:
\begin{align*}
U_m = \arg\min_{k \in S_m} \left| \phi_m^*  h_k\right|^2 
\end{align*}
where
\begin{align*}
S_m = \{ 1\leq k \leq n: &\left| \phi_m^* h_k\right|^2 \leq \epsilon_n, \forall r \neq m; \\
& \left|g_{kj}\right|^2 \leq \epsilon_n, \forall j \in \mathcal{\bar T}\}
\end{align*}
for the same $\epsilon_n$ sequence as in the downlink, where $\phi_m$ is the $m$th column of $\Phi$, \emph{i.e.}, the candidate set of users for stream $m$ are the users who receive bounded uplink interference as well as bounded inter-stream interference. We denote the set of uplink users scheduled in this way as $\mathcal{T} = \{ U_m\}_{m=1}^M$. 

\begin{remark}
Originally, random beamforming was considered for downlink communication in order to artifically induce channel variations and realize the multiuser diversity effect \cite{ViswanathTse_02}. However, in a full-duplex system, one also needs to induce variations in the level of interference to each user to extract this gain. Since each user has a single antenna, this is not possible through random beamforming at the uplink user side. However, one can still perform receive beamforming for uplink at the base station, which results in scheduling a different subset of users at each time slot, which in turn causes variations in the aggregate interference strength observed at each downlink user, as desired.
\end{remark}
\begin{remark}
Note that the base station or the uplink users do not require the channel knowledge of the interfering links for this scheme to work. If the downlink users are able to track the uplink interference strength they receive (which can potentially be arranged by overhearing the uplink pilots), they can send $\SNR$ feedback for their own channels only if the current interference level is below the threshold, and the base station can perform scheduling based only on this information.
\end{remark}

\subsection{Asymptotic Sum Capacity for Fixed Number of Antennas}

Define the achieved uplink and downlink gaps from individual uplink and downlink capacities as
\begin{align*}
\bar \eta\lp \mathcal{H}_n\rp &:= \mathbf{\bar C}^{\MAC}_n\lp \mathcal{H}_n\rp - \mathbf{\bar R}_n \lp \mathcal{H}_n\rp \\
\eta\lp \mathcal{H}_n\rp &:= \mathbf{C}^{\BC}_n\lp \mathcal{H}_n\rp - \mathbf{R}_n \lp \mathcal{H}_n\rp
\end{align*}
respectively, where $\mathbf{\bar C}^{\MAC}_n\lp \mathcal{H}_n\rp$ is the sum capacity of the multi-antenna MAC formed by considering the isolated uplink system, subject to the constraint that only $M$ users can transmit simultaneously, and $\mathbf{ C}^{\BC}_n\lp \mathcal{H}_n\rp$ is the sum capacity of the multi-antenna broadcast channel formed by isolating downlink system, achieved by dirty-paper coding \cite{WeingartenSteinberg_06}.

Clearly, $\mathbf{\bar C}^{\MAC}_n\lp \mathcal{H}_n\rp + \mathbf{C}^{\BC}_n\lp \mathcal{H}_n\rp$ is an upper bound on the sum rate $\mathbf{R}_n \lp \mathcal{H}_n\rp + \mathbf{\bar R}_n \lp \mathcal{H}_n\rp$ achievable in the full-duplex system. Our main result is that in a homogeneous network, this upper bound is asymptotically achievable as the number of users $n$ goes to infinity. This is more precisely stated in the following theorem.

\begin{theorem}\label{th:homo}
For any $\delta > 0$, 
\begin{align*}
\lim_{n\to\infty} \Prob{ \bar \eta\lp \mathcal{H}_n\rp+ \eta\lp \mathcal{H}_n\rp > \delta} = 0
\end{align*}
\end{theorem}
\begin{IEEEproof}
See Appendix~\ref{ap:proof_homo}.
\end{IEEEproof}

Theorem~\ref{th:homo} implies that for a homogeneous network with sufficiently many users, the uplink-to-downlink interference can be mitigated through proper user scheduling to the extent that the uplink and downlink systems gets asymptotically decoupled. The main idea underlying this result is to exploit \emph{multiuser diversity}, in terms of both the richness in the channel vectors to and from the base station, and richness in the strength of the interfering link.

Another important point in Theorem~\ref{th:homo} is that not only does the sum rate has the same scaling law as the decoupled system (which scales as $M\log \log n$ for both uplink and downlink, as in the isolated uplink and downlink systems \cite{SharifHassibi_05}), but the \emph{additive} gap between the decoupled system sum capacity and the achievable full-duplex sum rate goes to zero. A similar behavior has been observed before for MIMO broadcast channels, where it has been shown that the achievable rate difference between zero-forcing beamforming and dirty paper coding goes to zero as $n\to\infty$ \cite{BayestehKhandani_08}. Our result shows that through random beamforming, the same result can be obtained for simultaneous uplink and downlink, in the presence of uplink-to-downlink interference.

\subsection{Scaling the Number of Antennas}

An important assumption in Theorem~\ref{th:homo} is that the number of antennas remain fixed as $n$ grows. This is a crucial assumption, since as $M$ grows, one would need to schedule a growing number of users simultaneously in order to realize the full multiplexing gain of the system, which would result in increasing uplink-to-downlink and inter-stream interference. Hence, an important question is whether a similar result would hold in the case where $M$ is scaling. In \cite{SharifHassibi_05}, it is shown that for an isolated downlink system, the spatial multiplexing gain can be preserved if $M$ is scaled like $O\lp \log n\rp$. Here, we show a similar result for the full-duplex system, which is given in the following theorem.
\begin{theorem}\label{th:antennascaling}
If $\lim_{n \to \infty} \frac{M}{\log n}=\alpha$ for some $\alpha>0$,
\begin{align*}
\lim_{n\to\infty} \frac{\mathbf{\bar R}_n \lp \mathcal{H}_n \rp + \mathbf{ R}_n\lp \mathcal{H}_n \rp}{2M}=\beta
\end{align*}
for some $\beta>0$, almost surely.
\end{theorem}
\begin{IEEEproof}
See Appendix~\ref{ap:antennascaling}.
\end{IEEEproof}
Hence, even when the number of antennas grows to support the large number of users, the full sum degrees of freedom of the system can still be fully utilized despite the growing level of uplink interference, provided that the number of antennas does not scale faster than logarithmically in $n$.

%% file: Hetero.tex
The main idea underlying the result in Theorem~\ref{th:homo} was to exploit the channel richness in the network to asymptotically decouple the uplink and downlink transmissions. We have seen that the homogeneous model described in Section~\ref{sec:model} provides sufficient richness for this purpose. However, such homogeneity may not present in an actual network. Instead, users may be densely clustered in certain areas, and sparsely located in others. In such a scenario, it may not be possible to simultaneously approach uplink and downlink sum capacities, since the lack of channel diversity might force one to schedule an uplink-downlink user pair with significant interference in between.

In order to study this opposite regime, we consider a specific class of clustered networks that takes such non-homogeneity to the extreme, and prove that it is not possible to achieve the sum capacity of the decoupled system in such networks. Although the model of networks that we consider is rather specific, the main insight derived from this model might apply to more general heterogeneous networks. 

\begin{figure}
\centering
\includegraphics{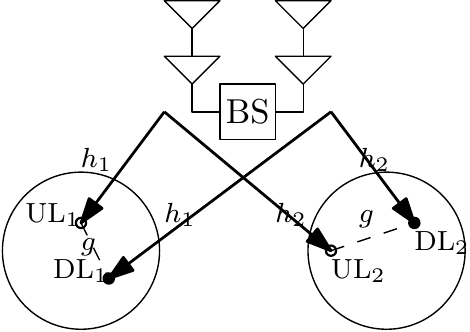}
\label{fig:homo}
\caption{A heterogeneous cellular system with a full-duplex base station with $M=2$ clusters and $n=2$ uplink and downlink half-duplex users. The uplink and downlink users within the same cluster have the same channel, and users in different clusters do not interfere with each other.}
\end{figure}

\subsection{Heterogeneous Model}
We consider a network with $M$ clusters (see Figure 2), hosting a total of $n$ uplink and $n$ downlink users that are uniformly distributed among them. We consider a simplified model where each cluster is assigned a spatial direction $h_i$, with $h_i^*h_j = 0$ for $i\neq j$, and $\|h_i\| = h$ for all $1\leq i \leq M$. We assume that all users (both uplink and downlink) within a cluster has the identical channel vector $h_i$. Further, we assume an all-or-none interference model, \emph{i.e.}, if $\kappa(i)$ denotes the cluster index of user $i$, then the interference link gain magnitude from user $j$ to user $i$ is given by
\begin{align*}
\left| G_{ij} \right| = \left\{ 
\begin{array}{ll}
g, & \text{ if $\kappa(i)=\kappa(j)$} \\
0, & \text{otherwise}
\end{array}
\right.
\end{align*}
As in the homogeneous case, we impose the constraint that at most $M$ uplink users can transmit simultaneously. Although this is a very simplified model, the unusual way in which the multi-antenna MAC and the BC interact with each other still makes this a non-trivial problem.

Henceforth, this model will be referred to as a $\lp M,h,g\rp$-clustered network. Next, we derive an upper bound on the sum capacity of the network. 

\subsection{Sum Rate Upper Bound and the Gap from the Decoupled System Capacity}
\begin{theorem}\label{th:hetero}
If $\lp \mathbf{\bar R}_n , \mathbf{R}_n\rp$ is an achievable rate pair in a $\lp M, h,g\rp-$clustered network, then
\begin{align*}
&\mathbf{\bar R}_n + \mathbf{R}_n < M\log\lp 1+\frac{h^2 \bar P}{1+g^2 \bar P}\rp \\
&\quad + M\log\lp 1 + h^2\frac{P}{M} + g^2\bar P + 2gh\sqrt{\frac{\bar P P}{M}}\rp 
\end{align*}
\end{theorem}
\begin{IEEEproof}
Let $y^{(m)}$ and $z^{(m)}$ denote the vector of channel outputs and the vector of noise at the users in cluster $m$. Since the downlink users do not cooperate, the capacity does not depend on the covariance matrix $\Sigma_z$ of the noise at the downlink, as long as $\Sigma_z \geq 0$ and the diagonal consists of $1$'s \cite{Sato_78}. Hence, we assume that within the same cluster, all downlink users are subject to the same noise process, \emph{i.e.}, $z_t^{(m)} \sim \mathcal{CN}(\mathbf{0}, \mathbf{1}\mathbf{1}^T)$, where $\mathbf{1}$ is the all ones vector. The noise processes at different clusters are independent\footnote{For a general broadcast channel, it is known that assuming independent noise processes gives a loose bound while using Sato upper bound \cite{ViswanathTse_03}; however, this does not matter in this case, since the links are orthogonal.}. Under these assumptions, using a genie-aided argument we show in Appendix~\ref{ap:upperbound} that for a block length of $N$,
\begin{align}
&N\lp \mathbf{\bar R}_n + \mathbf{R}_n\rp < \max_{\sum_{m=1}^M k_{m,t} \leq M} \;\; \max_{\frac{1}{N}\sum_{(m,t) \in \mathcal{N}} P_{m,t} \leq P} \notag\\
&\quad\sum_{t=1}^N \sum_{m=1}^M \log \lp 1 + h^2 P_{m,t} + k_{m,t} g^2 \bar P + 2gh\sqrt{k_{m,t}P_{m,t} \bar P}\rp  \notag \\&\quad+\log\lp 1+\frac{k_{m,t} h^2 \bar P}{1+k_{m,t} g^2 \bar P}\rp, \label{eq:upperbound}
\end{align}
where $k_{m,t}$ is the number of uplink users scheduled from cluster $m$ at time $t$, $P_{m,t}$ is the power allocated to $m$th channel at the base station at time $t$, and $\mathcal{N} := \lb M\rb \times\lb N\rb$. It can be verified that the $\log$ terms in \eqref{eq:upperbound} are concave and monotonically increasing in $\lp P_{m,t}, k_{m,t}\rp$, and hence the result follows by Jensen's inequality.
\end{IEEEproof}
It is easy to see that the sum of the isolated uplink and downlink capacities for a $\lp M,h,g\rp$-clustered network is given by
\begin{align} 
\bar C^{\MAC} + C^{\BC} = M\log\lp 1+h^2\frac{P}{M}\rp + M\log\lp 1+ h^2 \bar P\rp \label{eq:isolated}
\end{align}
Define the gaps form isolated systems, $\eta$ and $\bar \eta$ as in the homogeneous case. Also set $\SNR := h^2 \frac{P}{M}$, $\SNR^\alpha := g^2 \bar P$, $\SNR^\beta := h^2 \bar P$. The following corollary of Theorem~\ref{th:hetero} characterizes the scaling behavior of $\eta+\bar \eta$.
\begin{corollary} \label{cor:gap}
For a $\lp M,h,g\rp$-clustered network with number of users $n \geq M$,
\begin{align*}
\lim_{\SNR\to\infty} \frac{\eta+\bar \eta}{M\log \SNR} \geq 1
\end{align*}
\end{corollary}
\begin{IEEEproof}
See Appendix~\ref{ap:gap}.
\end{IEEEproof}
Note that this is the gap between an \emph{upper bound} on the sum capacity and the decoupled system capacity. Hence, regardless of the scheme applied, the achieved sum rate can get arbitrarily far from the decoupled system sum capacity.

\subsection{Potential for Cooperation over Side-Channels}
In order to remedy this inherent limitation in heterogeneous networks, we propose the use of device-to-device side-channels for user cooperation to resolve the full-duplex interference. In particular, we consider a system architecture where each uplink user is capable of allocating some $\lambda \in \lb 0,1\rb$ fraction of its power to an orthogonal channel that is used for cooperation with the downlink users. The side-channels are modeled by
\begin{align*}
\wtild y_i = g\wtild x_j + \wtild z_i
\end{align*}
with the power constraint $\mathbb{E}\left|\wtild X_j\right|^2\leq \lambda \bar P$, for each uplink user $j$ and downlink user $i$ such that $\kappa(i)=\kappa(j)$, with $\wtild z_i \sim \mathcal{CN}(0.1)$. Hence, the side-channels can be considered as orthogonal broadcast channels for each uplink user (we assume each broadcast channel operates over a different band, hence they do not interfere).

It is easy to see that cooperation over such orthogonal side-channels can help mitigate the device-to-device interference. Some schemes have been proposed in \cite{BaiSabharwal_13} regarding how to use such side-channels. Here, we focus on the following very simple scheme as an example to demonstrate that side-channels can indeed be effective in mitigating full-duplex interference in clustered networks.

Each uplink user $j$ replicates its symbol over the main channel on the side-channel, with equal power allocation, \emph{i.e.}, $\wtild x_j = \bar x_j$, and $\lambda = \frac{1}{2}$. Each downlink user $i$ subtracts the output received over the side-channel $\wtild y_i$ from its output in the main channel $y_i$ to obtain
\begin{align*}
y_i - \wtild y_i = h_i x + z_i - \wtild z_i
\end{align*}
Note that as a result, the effective channels of each uplink and downlink gets isolated, but the signal-to-noise ratio gets halved for both uplink and downlink due to power allocation and noise superposition, respectively. Therefore, this scheme can achieve
\begin{align*}
\mathbf{\bar R}_n + \mathbf{R}_n < M\log\lp 1+h^2\frac{P}{2M}\rp + M\log\lp 1+ h^2 \frac{\bar P}{2}\rp
\end{align*}
which is easily seen to be within $2M$ bits of the isolated system capacity with the side-channels (since the side-channel cannot increase capacity in the isolated case \cite{BaiSabharwal_13}), independent of $\SNR$.

%% file: AppendixHomoPf.tex
Assume the $M$ streams are decoded in the order $\lp 1,\dots,M\rp$ for both uplink and downlink, and denote the rate achieved on the $m$th uplink (downlink) stream by $\mathbf{\bar R}_n^{(m)}$ $\lp \mathbf{\bar R}_n^{(m)}\rp$, with $\sum_{m=1}^M \mathbf{\bar R}_n^{(m)} = \mathbf{\bar R}_n$ and $\sum_{m=1}^M \mathbf{ R}_n^{(m)} = \mathbf{ R}_n$. Note that these rates are all random variables due to their dependence on $\mathcal{H}_n$, $\Phi$ and $\bar \Phi$, but in this proof we will suppress this dependence for brevity.

Define $p_n:=\Prob{k \in S_m}$ and $\bar p_n:=\Prob{k \in \bar S_m}$ for an arbitrary user $1\leq k \leq n$ and arbitrary $1\leq m \leq M$. Note that $p_n, \bar p_n \to 0$ as $n\to\infty$.

Define $\delta'_n= \frac{p_n}{c}$ for a large constant $c>0$, define $q_n := p_n - \delta'_n$ and $\bar q_n := \bar p_n - \bar \delta'_n$, and the events
\begin{align*}
\mathcal{\bar F}_m &:= \lbp \left|\bar S_m\right| \geq n \bar q_n \rbp &\mathcal{\bar G}_m &:= \lbp \max_{k\in \bar S_m} \left| \bar \phi_m^* \bar h_k\right|^2 > \epsilon_n \rbp\\
\mathcal{ F}_m &:= \lbp \left| S_m\right| \geq n q_n\rbp  &\mathcal{ G}_m &:= \lbp \max_{k\in S_m} \left| \phi_m^* h_k\right|^2 > \epsilon_n \rbp
\end{align*}
Let us choose $\epsilon_n = O\lp \frac{1}{\log n}\rp$. Then 
\begin{align}
&\Prob{ \bar \eta+ \eta > \delta}\overset{\aaaa}{\leq} \Prob{ \bar \eta> \frac{\delta}{2}}  + \Prob{ \eta > \frac{\delta}{2}} \notag\\
& \overset{\bbbb}{\leq} \sum_{m=1}^M \Prob{ \frac{1}{M}\mathbf{\bar R}^{\MAC}_n- \mathbf{\bar R}_n^{(m)}  > \frac{\delta}{2M}}\notag \\
&\quad  + \sum_{m=1}^M \Prob{  \frac{1}{M}\mathbf{\bar R}^{\BC}_n - \mathbf{R}_n^{(m)}> \frac{\delta}{2M}} \notag\\
& \overset{\cccc}{=} M \Prob{ \frac{1}{M}\mathbf{\bar R}^{\MAC}_n - \mathbf{\bar R}_n^{(m)}  > \frac{\delta}{2M}}\notag \\
&\qquad  +M \Prob{  \frac{1}{M}\mathbf{\bar R}^{\BC}_n - \mathbf{R}_n^{(m)} > \frac{\delta}{2M}} \notag\\
& \overset{\dddd}{\leq} M\Prob{ \left. \frac{1}{M}\mathbf{\bar R}^{\MAC}_n - \mathbf{\bar R}_n^{(m)}  > \frac{\delta}{2M} \right| \mathcal{\bar F}_m, \mathcal{\bar G}_m} \label{eq:bqn_ub_u}\\
&\quad + M\Prob{\left. \mathcal{\bar G}_m^c \right| \mathcal{\bar F}_m} + M\Prob{\mathcal{\bar F}_m^c}\label{eq:bqn_ub_u2} \\
&\quad + M\Prob{ \left.  \frac{1}{M}\mathbf{ R}^{\BC}_n - \mathbf{R}_n^{(m)}  > \frac{\delta}{2M} \right| \mathcal{F}_m, \mathcal{G}_m}\label{eq:bqn_ub} \\
&\quad + M\Prob{\left. \mathcal{G}_m^c \right| \mathcal{F}_m} +M \Prob{\mathcal{F}_m^c}
 \label{eq:bqn_ub2}
\end{align}
where (a) and (b) follow by the fact that $\sum_{k=1}^K a_k>x \Rightarrow \bigvee_{k=1}^K \lp a_k>x/K \rp$ and by union bound; (c) follows because uniformly random selection of $\bar \Phi$ and $\Phi$ from the space of unitary matrices induces exchangeable distributions $p\lp \bar\phi_1,\dots,\bar \phi_M\rp$ and $p\lp \phi_1,\dots, \phi_M\rp$ on their respective columns; and (d) follows by the law of total probability and by upper bounding probabilities by one. Of the remaining terms, we will focus only on \eqref{eq:bqn_ub} and \eqref{eq:bqn_ub2} here, to avoid repetition. The uplink counterparts of these terms, given in \eqref{eq:bqn_ub_u} and \eqref{eq:bqn_ub_u2}, are bounded in exactly the same way in what follows, except where noted.

First consider \eqref{eq:bqn_ub}. Note that the conditioning on $\mathcal{G}_m$ implies that $k^* \notin S_r$ for $r \neq m$, where $k^*$ is the strongest user in $S_m, \emph{i.e.}$, $k^* = \arg\max_{k\in S_m} \left| \phi^*_m h_k\right|^2$. This ensures that the user that is scheduled for stream $m$ is not already scheduled for another stream, and hence, using independent Gaussian codebooks and allocating equal power for each downlink stream, 
\begin{align}
\mathbf{ R}_n^{(m)} \geq \log\lp 1 + \frac{P}{M}\frac{\max_{k\in  S_m} \left|  \phi_m^*  h_k\right|^2}{1+(2M-1)\epsilon_n}\rp, \label{eq:gaussian_lb}
\end{align}
almost surely. Therefore,
\begin{align}
&\Prob{ \left.  \frac{1}{M}\mathbf{ R}^{DPC}_n - \mathbf{R}_n^{(m)} > \frac{\delta}{2M} \right| \mathcal{F}_m, \mathcal{G}_m} \notag\\
& \overset{\aaaa}{\leq} \Prob{ \left. \log\lp \frac{ 1 + \frac{P}{M} \max_{1\leq k\leq n} \| h_k\|^2 }{ 1+\frac{P}{M} \frac{\max_{k\in S_m} \left| \phi_m^* h_k\right|^2}{1+(2M-1)\epsilon_n}}\rp > \frac{\delta}{2M} \right| \mathcal{F}_m, \mathcal{G}_m} \notag\\ 
& \overset{\bbbb}{\leq} \Prob{ \left.  \frac{\max_{1\leq k\leq n} \| h_k\|^2}{\max_{k\in S_m} \left| \phi_m^* h_k\right|^2}  > \frac{1+\frac{\delta}{2M}}{1+(2M-1)\epsilon_n}  \right| \mathcal{F}_m, \mathcal{G}_m} \notag\\ 
& \overset{\cccc}{\leq} \mathbb{P}\lp \frac{\max_{1\leq k\leq n} \| h_k\|^2}{\max_{k \in S_m} \| h_k\|^2}+\frac{\max_{k \in S_m} \| h_k\|^2}{\max_{k\in S_m} \left| \phi_m^* h_k\right|^2}  \right.\notag \\
&\qquad\qquad \qquad\qquad \left.> \left.  2\sqrt{\frac{1+\frac{\delta}{2M}}{1+(2M-1)\epsilon_n}}  \right| \mathcal{F}_m, \mathcal{G}_m\rp \notag\\ 
& \overset{\dddd}{\leq} \Prob{ \left.  \frac{\max_{1\leq k\leq n} \| h_k\|^2}{\max_{k \in S_m} \| h_k\|^2}  > 1+\gamma \right| \mathcal{F}_m, \mathcal{G}_m} \label{eq:term1}\\
&\quad +\Prob{ \left. \frac{\max_{k \in S_m} \| h_k\|^2}{\max_{k\in S_m} \left| \phi_m^* h_k\right|^2}  > 1+\gamma  \right| \mathcal{F}_m, \mathcal{G}_m}   \label{eq:term2}
\end{align}
where (a) follows by using Lemma 3 in \cite{SharifHassibi_07} for downlink and Lemma~\ref{lem:sdma_ub} for uplink (replace $\frac{P}{M}$ with $\bar P$ for uplink); (b) follows by the inequality $e^x \geq 1 + x$ and by the fact that $\frac{x}{y}\geq\frac{1+x}{1+y}$ for $x\geq y$; (c) follows by the fact $ab\geq x \Rightarrow a+b\geq 2\sqrt{x}$ (by AM-GM inequality); (d) follows by the fact that $\sum_{k=1}^K a_k>x \Rightarrow \bigvee_{k=1}^K \lp a_k>x/K \rp$, by the union bound, and by defining $\gamma>0$ such that
\begin{align*}
\lp 1+(2M-1)\epsilon_n\rp\lp 1+ \gamma \rp^2 < 1+\frac{\delta}{2M}
\end{align*}
for sufficiently large $n$.

Next, we bound the terms \eqref{eq:term1} and \eqref{eq:term2} separately. Consider \eqref{eq:term1} first.
\begin{align*}
&\eqref{eq:term1}\leq\Prob{ \left.  \frac{\max_{1\leq k\leq n} \| h_k\|^2}{\max_{k \in S_m} \| h_k\|^2}  > 1 + \gamma  \right| \mathcal{F}_m, \mathcal{G}_m} \\
&\overset{\aaaa}{=} \Prob{ \left.  \frac{ \max_{k\in S_m^c} \| h_k\|^2}{\max_{k \in S_m} \| h_k\|^2}  > 1 + \gamma  \right| \mathcal{F}_m, \mathcal{G}_m}\\
&\leq \Prob{ \left.  \frac{ \max_{k\in S_m^c} \| h_k\|^2}{\max_{k \in S_m} \left| \phi_m^* h_k\right|^2}  > 1 + \gamma  \right| \mathcal{F}_m, \mathcal{G}_m}\\
&\overset{\bbbb}{\leq} \frac{1}{1-\epsilon'_n}\Prob{ \left.  \frac{ \max_{k\in S_m^c} \| h_k\|^2}{\max_{k \in S_m} \left| \phi_m^* h_k\right|^2}  > 1 + \gamma  \right|\left| S_m\right| > nq_n}\\
&= \frac{1}{1-\epsilon'_n}\sum_{s=\lceil nq_n \rceil}^n 
\sum_{\mathcal{ A}_s\subseteq \lb n\rb: \left| \mathcal{ A}_s\right|=s}
\Prob{\left. S_m = \mathcal{A}_s\right|\left| S_m\right| \geq nq_n} \\
& \cdot\Prob{ \left.  \frac{ \max_{k\in \mathcal{A}_s^c} \| h_k\|^2}{\max_{k \in \mathcal{A}_s} \left| \phi_m^* h_k\right|^2}  > 1 + \gamma  \right| \left| \mathcal{A}_s\right| \geq nq_n, S_m = \mathcal{A}_s}\\
&\overset{\cccc}{=} \frac{1}{1-\epsilon'_n}\sum_{s=\lceil nq_n \rceil}^n 
\sum_{\mathcal{ A}_s\subseteq \lb n\rb: \left| \mathcal{ A}_s\right|=s}
\Prob{\left. S_m = \mathcal{A}_s\right|\left| S_m\right| \geq nq_n} \\
& \cdot\Prob{ \left.  \frac{ \max_{k\in \mathcal{A}_s^c} \| h_k\|^2}{\max_{k \in \mathcal{A}_s} \left| \phi_m^* h_k\right|^2}  > 1 + \gamma  \right| \left| \mathcal{A}_s\right| \geq nq_n, S_m = \mathcal{A}_s ,\right.\\
&\left. \vphantom{ \frac{ \max_{k\in \mathcal{A}_s^c} \| h_k\|^2}{\max_{k \in \mathcal{A}_s} \left| \phi_m^* h_k\right|^2} }
 \lbp \exists r\neq m:\left| \phi^*_r h_{k^*}\right|^2 > \epsilon_n\; \vee\; \exists j \in \mathcal{\bar T}: \left| h_{k^*j}\right|^2>\epsilon_n\rbp}\\
&\overset{\dddd}{=}\frac{1}{1-\epsilon'_n}\sum_{s=\lceil nq_n \rceil}^n 
\Prob{\left. \left|S_m\right| = s\right|\left| S_m\right| > nq_n } \\
& \cdot\Prob{ \left.  \frac{ \max_{k\in \mathcal{A}_s^c} \| h_k\|^2}{\max_{k \in \mathcal{A}_s} \left| \phi_m^* h_k\right|^2}  > 1 + \gamma  \right|\left| \mathcal{A}_s\right| = s, \right. \\
& \left. \vphantom{\frac{ \max_{k\in \mathcal{A}_s^c} \| h_k\|^2}{\max_{k \in \mathcal{A}_s} \left| \phi_m^* h_k\right|^2}}
\lbp \exists r\neq m:\left| \phi^*_r h_{k^*}\right|^2 > \epsilon_n\; \vee\; \exists j \in \mathcal{\bar T}: \left| h_{k^*j}\right|^2>\epsilon_n\rbp}\\
&\overset{\eeee}{\leq}\frac{1}{1-\epsilon'_n}\Prob{ \left.  \frac{ \max_{k\in \mathcal{\tilde A}_{s}^c} \| h_k\|^2}{\max_{k \in \mathcal{\tilde A}_{s}} \left| \phi_m^* h_k\right|^2}  > 1 + \gamma  \right| \right. \\
& \left.\vphantom{\frac{ \max_{k\in \mathcal{\tilde A}^c} \| h_k\|^2}{\max_{k \in \mathcal{\tilde A}} \left| \phi_m^* h_k\right|^2}}
\lbp \exists r\neq m:\left| \phi^*_r h_{k^*}\right|^2 > \epsilon_n\; \vee\; \exists j \in \mathcal{\bar T}: \left| h_{k^*j}\right|^2>\epsilon_n\rbp}\\
&\overset{\ffff}{\leq}\frac{1}{\lp1-\epsilon'_n\rp^2}\Prob{ \frac{ \max_{k\in \mathcal{\tilde A}^c} \| h_k\|^2}{\max_{k \in \mathcal{\tilde A}} \left| \phi_m^* h_k\right|^2}  > 1 + \gamma  }\\
&\overset{\gggg}{\leq}\frac{1}{\lp1-\epsilon'_n\rp^2} \lb \Prob{\max_{k \in \mathcal{\tilde A}} \left| \phi_m^* h_k\right|^2 < 2\log\lp \frac{n q_n}{\log\lp nq_n\rp}\rp} \right.\\
& + \mathbb{P}\lp\max_{k\in \mathcal{\tilde A}^c} \| h_k\|^2 >2\log\lp n q_n\rp +(2M+2)\log\log\lp nq_n\rp \rp \\
& + \left. \mathbb{P}\lp  \frac{2\log\lp nq_n\rp +(2M+2)\log\log\lp nq_n\rp}{2\log\lp n q_n\rp -2\log\log\lp nq_n\rp}  > 1 + \gamma \rp \rb \\
&\overset{\hhhh}{\leq} \frac{1}{\lp1-\epsilon'_n\rp^2}\lb \frac{1}{nq_n} + O\lp \frac{1}{\log\lp nq_n\rp}\rp + 0 \rb\\
&\overset{\iiii}{=} O\lp \frac{1}{\log n}\rp
\end{align*}
where
\begin{itemize}
\item (a) follows by the fact that the ratio can be larger than $\lp1+\gamma\rp$ only if the maximum in the numerator occurs for a $k\in S_m^c$ (otherwise the ratio is 1);
\item (b) is by the fact that for events $A,B$; $\Prob{A|B}\leq\frac{\Prob{A}}{\Prob{B}}$ and by Lemma~\ref{lem:Gn_epsilon}, where $\epsilon_n' \to 0$;
\item (c) is because $S_m = \mathcal{A}_s$ implies the newly conditioned event, which is that for any user outside the set $\mathcal{A}_s$, there must exist an $r$ such that $\left| \phi^*_r h_{k^*}\right|^2>\epsilon_n$ or an uplink user whose interference strength is larger than $\epsilon_n$, by the construction of the set $S_m$, where we have defined $k^* := \arg\max_{k \in \mathcal{A}_s^c}\| h_k\|^2$ (for the uplink case the second part of the event is removed);
\item (d) follows by the fact that the probability on the right-hand side does not depend on $\mathcal{A}_s$, as long as $\left| \mathcal{A}_s\right|$ is fixed, owing to the fact that the user channel vectors $\bar h_k$ are i.i.d.;
\item (e) follows because the given probability is a monotonically decreasing function of $s$, and $\mathcal{\tilde A}$ is any arbitrary subset of users such that $\left| \mathcal{\tilde A}\right| = \lceil nq_n\rceil$;
\item (f) is by Lemma~\ref{lem:Gn_epsilon};
\item (g) is by the fact that for events $A,B,C$; $\Prob{A} \leq \Prob{B^c} + \Prob{C^c} + \Prob{A|B,C}$ by union bound and law of total probability;
\item (h) is because of Lemmas~\ref{lem:chi2} and \ref{lem:chi2m}, and by the fact that the last probability is that of the elements of a deterministic sequence converging to 1 being larger than $1+\gamma$ for sufficiently large $n$;
\item (i) is because we chose $\epsilon_n=O\lp \frac{1}{\log n}\rp$, and thus $\bar q_n = O\lp \frac{1}{\log^{M-1} n}\rp$ by Lemma~\ref{lem:p_ul} for the uplink and $q_n = O\lp \frac{1}{\log^{2M-1} n}\rp$ by Lemma~\ref{lem:p_dl} for the downlink.
\end{itemize}
Next, we move on to analyze the term \eqref{eq:term2}.
\begin{align*}
&\eqref{eq:term2}\leq \Prob{ \left. \frac{\max_{k \in S_m} \| h_k\|^2}{\max_{k\in S_m} \left| \phi_m^* h_k\right|^2}  > 1+\gamma  \right| \mathcal{F}_m, \mathcal{G}_m}\\
&\overset{\aaaa}{=} \Prob{ \left. \frac{\max_{k \in S_m} \sum_{r=1}^M\left| \phi_r^*h_k\right|^2}{\max_{k\in S_m} \left| \phi_m^* h_k\right|^2}  > 1+\gamma  \right| \mathcal{F}_m, \mathcal{G}_m}\\
&\overset{\bbbb}{\leq} \Prob{ \left. \frac{\max_{k \in S_m} \left| \phi_m^*h_k\right|^2 + M\epsilon_n}{\max_{k\in S_m} \left| \phi_m^* h_k\right|^2}  > 1+\gamma  \right| \mathcal{F}_m, \mathcal{G}_m}\\
&= \Prob{ \left. \max_{k \in S_m} \left| \phi_m^*h_k\right|^2 <  \frac{M\epsilon_n}{\gamma}\right|  \mathcal{F}_m, \mathcal{G}_m}\\
&\overset{\cccc}{\leq} \frac{1}{1-\epsilon'_n}\Prob{ \left. \max_{k \in S_m} \left| \phi_m^*h_k\right|^2 <  \frac{M\epsilon_n}{\gamma}\right| \left| S_m\right| \geq nq_n}\\
&\overset{\dddd}{\leq} \frac{1}{1-\epsilon'_n} \lp 1 - \exp\lbp -\frac{M\epsilon_n}{2\gamma}\rbp\rp^{nq_n} \\
&\overset{\eeee}{=} O\lp \frac{1}{\lp\log n\rp^{n/\log n}}\rp
\end{align*}
where
\begin{itemize}
\item (a) follows by the fact that $\Phi$ is unitary and thus $\|\Phi h_k\|=\|h_k\|$; 
\item (b) is by construction of the set $S_m$;
\item (c) is by Lemma~\ref{lem:Gn_epsilon};
\item (d) is by Lemma~\ref{lem:main};
\item (e) is because we chose $\epsilon_n=O\lp \frac{1}{\log n}\rp$, thus $q_n = O\lp \frac{1}{\log^{2M-1} n}\rp$ by Lemma~\ref{lem:p_dl}, and $\bar q_n = O\lp \frac{1}{\log^{M-1} n}\rp$ by Lemma~\ref{lem:p_ul}.
\end{itemize}
Therefore \eqref{eq:bqn_ub} goes to zero as $n\to\infty$. Next, we consider the terms in \eqref{eq:bqn_ub2}. Note that the first term goes to zero since
\begin{align*}
\Prob{\mathcal{G}^c|\mathcal{F}} &= \lp 1 - \exp\lbp -\frac{\epsilon_n}{2}\rbp\rp^{nq_n} \\
&= O\lp \frac{1}{\lp\log n\rp^{n/\log n}}\rp
\end{align*}
by Lemma~\ref{lem:main} and by the choice of $\epsilon_n$. The second term in \eqref{eq:bqn_ub2} goes to zero by weak law of large numbers for triangular arrays \cite{Durrett_10}, applied to the binomial random variable $\left| S_m\right|$ with mean $np_n$.

Since all terms in \eqref{eq:bqn_ub_u}, \eqref{eq:bqn_ub_u2}, \eqref{eq:bqn_ub}, and \eqref{eq:bqn_ub2} go to zero, the result follows.

%% file: AppendixAntennaScaling.tex
Let us choose $\epsilon_n = \epsilon>0$, \emph{i.e.}, a constant. Then, as in the proof of Theorem~\ref{th:homo},
\begin{align*}
\Prob{\mathbf{R}_n + \mathbf{\bar R}_n < 2M\beta } &\leq \sum_{m=1}^M \Prob{\mathbf{R}_n^{(m)} < \beta} \\
&\qquad + \Prob{\mathbf{\bar R}_n^{(m)} < \beta} 
\end{align*}
by the fact that $\sum_{k=1}^K a_k<x \Rightarrow \bigvee_{k=1}^K \lp a_k<x/K \rp$ and by union bound.
We only consider the first term, associated with downlink. The uplink term is bounded the same way, except where noted. By law of total probability, and by upper bounding probabilities by one,
\begin{align}
&\Prob{\mathbf{R}_n^{(m)} < \beta} \leq \Prob{\mathcal{F}^c_m} + \Prob{\mathcal{G}_m^c|\mathcal{F}_m} \notag\\
&\qquad+\Prob{\left. \mathbf{R}_n^{(m)} < \beta \right| \mathcal{F}_m, \mathcal{G}_m} \label{eq:as_main}
\end{align}
Since $\epsilon_n$ is a constant, $\Prob{\mathcal{F}_m^c}$ goes to zero exponentially by Hoeffding's inequality. $\Prob{\mathcal{G}_m^c|\mathcal{F}_m} $ is upper bounded by
\begin{align*}
\Prob{\mathcal{G}_m^c|\mathcal{F}_m} &\leq \lp 1 - \exp\lbp -\frac{\epsilon}{2}\rbp \rp^{nq_n} \\
&=a^{\frac{n^{1+2\log a}}{a}},
\end{align*}
by Lemmas~\ref{lem:main} and \ref{lem:p_dl}, where $a = \lp 1-\exp\lbp -\epsilon/2\rbp\rp$. Note that the last term goes to zero exponentially if $1+2\log a >0$, which is satisfied for sufficiently large $\epsilon>0$.
We consider the first term. Conditioned on $\mathcal{G}_m$, a different user is scheduled for each stream, hence
\begin{align*}
&\Prob{\left.\mathbf{R}_n^{(m)} < \beta \right| \mathcal{F}_m, \mathcal{G}_m} \\
&\leq \Prob{\left.\log\lp \frac{\max_{k\in S_m} \left| \phi_m^* h_k\right|^2}{1+(2M-1)\epsilon}\rp < \beta\right| \mathcal{F}_m, \mathcal{G}_m}\\
&\overset{\aaaa}{=}\frac{1}{1-\epsilon_n'}\Prob{\left.\log\lp \frac{\max_{k\in S_m} \left| \phi_m^* h_k\right|^2}{1+(2M-1)\epsilon}\rp < \beta\right| \mathcal{F}_m}\\
&\leq\frac{1}{1-\epsilon_n'}\Prob{\left. \max_{k\in S_m} \left| \phi_m^* h_k\right|^2 < \beta e\lp 1+2\epsilon \log n\rp\right| \mathcal{F}_m} \\
&\overset{\bbbb}{\leq} \frac{1}{1-\epsilon_n'}\lp 1 - \exp\lbp -\frac{\beta e\lp 1+2\epsilon \log n\rp}{2}\rbp\rp^{nq_n}\\
&\overset{\cccc}{=} \Theta\lp e^{-n^\gamma}\rp
\end{align*}
where (a) follows by Lemma~\ref{lem:Gn_epsilon}, (b) follows by Lemma~\ref{lem:main}, and (c) follows, for some $0<\gamma<1$, by Lemma~\ref{lem:p_dl} with the choice $\epsilon_n =\epsilon$, and by letting $M=\alpha \log n$ for sufficiently small $\alpha>0$. Since all terms in \eqref{eq:as_main} go to zero exponentially as $n \to \infty$, 
\begin{align*}
\sum_{n} \Prob{\mathbf{R}_n + \mathbf{\bar R}_n < 2M } < \infty
\end{align*}
and thus by Borel-Cantelli Lemma \cite{Durrett_10}, the result follows.

%% file: AppendixUpperBound.tex
Let us denote message of the $k$th uplink user as $\bar W_k$, the message intended for the $k$th downlink user by $W_k$, and for any set $S$, define $W_S = \lbp W_k : k \in S\rbp$. We also define $v_t^{(m)} = y_t^{(m)} - \mathbf{1} h_m^*x_t$, where $\mathbf{1}$ is the vector  of ones, \emph{i.e.}, $v_t^{(m)}$ is the vector of interference signals at the downlink users of cluster $m$ at time $t$. Set $v_t = \lb v_t^{(1)},\dots,v_t^{(M)}\rb^*$.

We consider a block length of $N$, and as explained in Section~\ref{sec:hetero}, assume $z_t^{(m)} \sim \mathcal{CN}(\mathbf{0}, \mathbf{1}\mathbf{1}^T)$, where $\mathbf{1}$ is the all ones vector, for $m\in \lb M\rb$. We also assume that the downlink users within each cluster cooperate, since this cannot reduce capacity. Then, by Fano's inequality,
\begin{align*}
&N\lp \mathbf{R}_n+\mathbf{\bar R}_n\rp \leq I\lp W_{[n]} ; y^N\rp + I\lp \bar W_{[n]} ; \bar y^N\rp \\
&\leq  I\lp W_{[n]} ; y^N\rp + I\lp \bar W_{[n]} ; \bar y^N, y^N, W_{[n]}\rp \\
&\overset{\aaaa}{=}  I\lp W_{[n]} ; y^N\rp + I\lp \bar W_{[n]} ; \bar y^N, y^N \left| W_{[n]} \right.\rp \\
&=  h\lp  y^N\rp - h\lp y^N \left| W_{[n]} \right.\rp + h\lp \bar y^N, y^N \left| W_{[n]} \right.\rp \\
&\qquad - h\lp \bar y^N, y^N \left| W_{[n]}, \bar W_{[n]}  \right.\rp \\
&=  h\lp  y^N\rp + h\lp \bar y^N \left| W_{[n]}, y^N  \right.\rp - h\lp \bar y^N, y^N \left| W_{[n]}, \bar W_{[n]}  \right.\rp \\
&=  \sum_{t=1}^N h\lp  y_t | y^{t-1}\rp + h\lp \bar y_t \left| W_{[n]}, y^N, \bar y^{t-1}  \right.\rp \\
&\qquad - h\lp \bar y_t, y_t \left| W_{[n]}, \bar W_{[n]}, \bar y^{t-1}, y^{t-1}  \right.\rp \\
&\overset{\bbbb}{=}  \sum_{t=1}^N h\lp  y_t | y^{t-1}\rp + h\lp \bar y_t \left| W_{[n]}, y^N, \bar y^{t-1}, x_t  \right.\rp\\
&\qquad  - h\lp \bar y_t, y_t \left| W_{[n]}, \bar W_{[n]}, \bar y^{t-1}, y^{t-1}, \bar x_t, x_t  \right.\rp \\
&\overset{\cccc}{\leq}  \sum_{t=1}^N h\lp  y_t\rp + h\lp \bar y_t \left| y_t, x_t  \right.\rp - h\lp \bar z_t, z_t\rp \\
&\overset{\dddd}{=}  \sum_{t=1}^N h\lp  y_t\rp + h\lp \bar y_t \left| v_t, x_t  \right.\rp - h\lp \bar z_t\rp - h\lp z_t\rp \\
&\leq  \sum_{t=1}^N h\lp  y_t\rp + h\lp \bar y_t \left| v_t \right.\rp - h\lp \bar z_t\rp - h\lp z_t\rp \\
&\overset{\eeee}{\leq}  \sum_{t=1}^N \lp\sum_{m=1}^M h\lp  y_t^{(m)}\rp\rp + h\lp \bar y_t \left| v_t \right.\rp  - h\lp \bar z_t\rp\\
&\qquad - \lp\sum_{m=1}^M h\lp z_t^{(m)}\rp\rp
\end{align*}
where (a) follows by independence of messages; (b) follows by the fact that $x_t$ is a deterministic function of $\lp W_{[n]}, \bar y^{t-1}\rp$ and $\bar x_t$ is a deterministic function of $\bar W_{[n]}$; (c) follows because conditioning reduces entropy and by subtracting $x_t$ and $\bar x_t$ from $y_t$ and $\bar y_t$; (d) is because $v_t = y_t - \mathbf{1}H^* x$ and by independence of uplink and downlink noise; (e) is by the fact that conditioning reduces entropy, and that noise processes at different clusters are independent. Since $\lbp h_m\rbp$ are orthogonal, $\lbp \bar y_t^{(m)}\rbp$ can be uniquely expressed as $\bar y_t = \sum_{m=1}^M \frac{h^*_m}{\|h_m\|} \bar y_t^{(m)}$, \emph{i.e.}, this transformation is a bijection. Let us define the matrix $\wtild H := \lb \frac{h_1}{\|h_1\|} \; \dots\; \frac{h_M}{\|h_M\|}\rb$. Then
\begin{align*}
h\lp \bar y_t\rp &= h\lp \wtild H^* \bar y_t\rp = h\lp \bar y_t^{(1)},\dots,\bar y_t^{(M)}\rp + \log \left| \wtild H\right| \\
&= h\lp \bar y_t^{(1)},\dots,\bar y_t^{(M)}\rp
\end{align*}
since $\wtild H$ is unitary. Similarly, $\bar z_t = \sum_{m=1}^M h_m \bar z_t^{(m)}$, and $\lbp \bar z_t^{(m)}\rbp$ are still distributed i.i.d. $\mathcal{CN}(0,1)$. Hence, also using the fact that conditioning reduces entropy,
 \begin{align*}
N\lp \mathbf{R}_n+\mathbf{\bar R}_n\rp &\leq  \sum_{t=1}^N \sum_{m=1}^M h\lp  y_t^{(m)}\rp + h\lp \bar y_t^{(m)} \left| v_t^{(m)} \right.\rp \\
&\quad -  h\lp \bar z_t^{(m)}\rp -   h\lp z_t^{(m)}\rp
\end{align*}
Let $k_{m,t}$ denote the number of uplink users scheduled in cluster $m$ at time $t$, with $\sum_{m=1}^M k_{m,t} \leq M$, for all $t$. Note that given any power allocation, there is a covariance constraint on $\lb \begin{array}{cc} \bar y_t^{(m)} & v_{t}^{(m)} \end{array}\rb^*$ given by
\begin{align*}
K = I + k_{m,t}\bar P\lb \begin{array}{c} h_m \\ g \end{array}\rb\lb\begin{array}{cc} h_m^* & g* \end{array}\rb.
\end{align*}
Hence, $h(\bar y_t| v_t)$ is maximized when $(\bar y_t, v_t) \sim \mathcal{CN}\lp 0, K\rp$, with
\begin{align*}
h(\bar y_t| v_t) = \log 2\pi e \left| K_{\bar y|v}\right|,
\end{align*}
where $K_{\bar y|v}$ is the conditional covariance matrix of $\bar y_t^{(m)}$ given $v_{t}^{(m)}$. Therefore, evaluating the differential entropy terms with Gaussian input distributions\footnote{We evaluate $h(y_t^{(m)})$ assuming a \emph{joint} Gaussian distribution on $x_t$ and $\bar x_t$ with arbitrary correlation, since $x_t$ is a function of both $W_{[n]}$ and $\bar y_{t-1}$.}, and using the fact that $z_t^{(m)} \sim \mathcal{CN}(\mathbf{0}, \mathbf{1}\mathbf{1}^T)$, we find \eqref{eq:upperbound}.

%% file: AppendixHeteroGap.tex
Using Theorem~\ref{th:hetero} and \eqref{eq:isolated}, $\eta+\bar \eta$ can be lower bounded by
\begin{align*}
&\eta+\bar \eta > M\log\lp  \frac{1+\SNR^\beta}{1+\frac{\SNR^\beta}{1+\SNR^\alpha}}\rp \\
&\quad - M\log\lp 1 + \frac{\SNR^\alpha}{1 + \frac{1}{M}\SNR}\rp - M\log 3
\end{align*}
If we use the notation $f\lp \SNR \rp \doteq g\lp \SNR \rp$ to mean that $\lim_{\SNR\to\infty}\frac{f\lp \SNR\rp}{g\lp \SNR \rp}=1$, then it is easy to see that
\begin{align*}
&\log\lp  \frac{1+\SNR^\beta}{1+\frac{\SNR^\beta}{1+\SNR^\alpha}}\rp-\log\lp 1 + \frac{\SNR^\alpha}{1 + \frac{1}{M}\SNR}\rp -\log 3\\
&\doteq \log \SNR^\alpha - \log \SNR^{\alpha-1} \\
&= \log \SNR
\end{align*}
Hence, the result follows.

%% file: AppendixLemmas.tex
\begin{lemma}\label{lem:sdma_ub}
\begin{align*}
\mathbf{\bar R}^{\MAC}_n\lp \mathcal{H}_n\rp \leq M\log\lp 1 + P \max_{1\leq k \leq n} \| \bar h_k\|^2\rp
\end{align*}
\end{lemma}
\begin{IEEEproof}
The capacity of a MIMO MAC with a per-user power constraint $\bar P$, and an active user constraint $M$ is given by
\begin{align*}
\mathbf{\bar R}^{\MAC}_n\lp \mathcal{H}_n\rp &= \max_{\mathcal{A}\subseteq\lb n \rb: \left| \mathcal{A}\right|=M} \log \left| I_M + \bar P\bar H_{\mathcal{A}} \bar H_{\mathcal{A}}^* \right| \\
&= \max_{\mathcal{A}\subseteq\lb n \rb: \left| \mathcal{A}\right|=M} \log \left| I_M + \bar P\sum_{k \in \mathcal{A}} \bar h_k \bar h_k^* \right|
\end{align*}
Using the inequality $|A|\leq\lp\frac{\text{tr}(A)}{M}\rp^M$ (which is a direct consequence of AM-GM inequality applied to the eigenvalues of $A$), 
\begin{align*}
&\mathbf{\bar R}^{\MAC}_n\lp \mathcal{H}_n\rp \\
&= \max_{\mathcal{A}\subseteq\lb n \rb: \left| \mathcal{A}\right|=M} M\log \lp\frac{\text{tr}\lp I_M + \bar P \sum_{k \in \mathcal{A}} \bar h_k \bar h_k^* \rp}{M}\rp\\
&=\max_{\mathcal{A}\subseteq\lb n \rb: \left| \mathcal{A}\right|=M} M\log \lp 1 +\frac{\bar P\sum_{k \in \mathcal{A}}\text{tr}\lp  \bar h_k \bar h_k^* \rp}{M}\rp\\
&=\max_{\mathcal{A}\subseteq\lb n \rb: \left| \mathcal{A}\right|=M} M\log \lp 1 +\bar P\frac{\sum_{k \in \mathcal{A}}\| \bar h_k\|^2}{M}\rp\\
&= M\log \lp 1 +\bar P\max_{\mathcal{A}\subseteq\lb n \rb: \left| \mathcal{A}\right|=M}\frac{\sum_{k \in \mathcal{A}}\| \bar h_k\|^2}{M}\rp\\
&\leq M\log \lp 1 +\bar P \max_{1\leq k \leq n}\| \bar h_k\|^2\rp
\end{align*}
\end{IEEEproof}

\begin{lemma} \label{lem:p_ul}
For an arbitrary uplink user $1\leq k \leq n$, and arbitrary $1\leq m\leq M$,
\begin{align*}
\Prob{k\in \bar S_m} = \lp 1-\exp\lbp-\epsilon_n/2\rbp \rp^{M-1}
\end{align*}
\begin{IEEEproof}
\begin{align*}
\Prob{k\in \bar S_m} &= \Prob{\left| \bar \phi^*_r \bar h_k\right|^2 \leq \epsilon_n,\; \forall r \neq m} \\
&\overset{\aaaa}{=}\lb\Prob{\left| \bar \phi^*_1 \bar h_k\right|^2 \leq \epsilon_n} \rb^{M-1}\\
&\overset{\bbbb}{=}\lp 1-\exp\lbp-\epsilon_n/2\rbp \rp^{M-1}
\end{align*}
where (a) follows by the fact that the components of $\bar \Phi \bar h_k$ are i.i.d. distributed because $\bar \Phi$ is unitary; and (b) follows by the fact that $\left| \bar \phi^*_1 \bar h_k\right|^2$ is $\chi^2(2)$ distributed.
\end{IEEEproof}
\end{lemma}

\begin{lemma} \label{lem:p_dl}
For an arbitrary downlink user $1\leq k \leq n$, and arbitrary $1\leq m\leq M$,
\begin{align*}
\Prob{k\in S_m} = \lp 1-\exp\lbp-\epsilon_n/2\rbp \rp^{2M-1}
\end{align*}
\begin{IEEEproof}
\begin{align*}
&\Prob{k\in S_m} \\
&= \Prob{\left| \phi^*_r h_k\right|^2 \leq \epsilon_n,\; \forall r \neq m;\;\;\left|h_{kj}\right|^2\leq\epsilon_n,\;\forall j\in\mathcal{\bar T}} \\
&= \sum_{\mathcal{A}\subseteq \lb n \rb: \left| \mathcal{A}\right|=M}\Prob{\mathcal{\bar T}=\mathcal{A}}\\ &\cdot\Prob{\left.\left| \phi^*_r h_k\right|^2 \leq \epsilon_n \;\forall r \neq m;\;\;\left|h_{kj}\right|^2\leq\epsilon_n,\;\forall j\in\mathcal{A}\right| \mathcal{\bar T}=\mathcal{A}} \\
&\overset{\aaaa}{=} \sum_{\mathcal{A}\subseteq \lb n \rb: \left| \mathcal{A}\right|=M}\Prob{\mathcal{\bar T}=\mathcal{A}} \\
&\cdot\Prob{\left| \phi^*_r h_k\right|^2 \leq \epsilon_n \;\forall r \neq m;\;\;\left|h_{kj}\right|^2\leq\epsilon_n,\;\forall j\in\mathcal{A}} \\
&= \Prob{\left| \phi^*_r h_k\right|^2 \leq \epsilon_n \;\forall r \neq m;\;\;\left|h_{kj}\right|^2\leq\epsilon_n,\;\forall j\in\mathcal{A}} \\
&\overset{\bbbb}{=} \lb\Prob{\left| \bar \phi^*_r \bar h_k\right|^2 \leq \epsilon_n} \rb^{M-1} \lb\Prob{\left|h_{k1}\right|^2\leq\epsilon} \rb^M \\
&\overset{\cccc}{=}\lp 1-\exp\lbp-\epsilon_n/2\rbp \rp^{2M-1}
\end{align*}
where (a) follows by the fact that $\mathcal{\bar T}$ is a function of $\lbp \bar \phi^*_m \bar h_k\rbp_{m,k}$, and all links are independent, and thus the event $\lbp \mathcal{\bar T} = \mathcal{A}\rbp$ is independent, (defining $\mathcal{\tilde A}$ to be an arbitrary subset of uplink users s.t. $\left| \mathcal{\tilde A}\right|=M$); (b) follows because the components of $ \Phi  h_k$ are i.i.d. distributed and all links are independent; and (c) follows because both $\left| \phi^*_1 h_k\right|^2$ and $\left| h_{k1}\right|^2$ are $\chi^2(2)$ distributed.
\end{IEEEproof}
\end{lemma}

\begin{lemma}\label{lem:main}
\begin{align*}
\Prob{ \left. \max_{k \in S_m} \left| \phi_m^*h_k\right|^2 <  x\right| \left| S_m\right| \geq nq_n} \leq \lp  1-e^{ -\frac{x}{2}}\rp^{nq_n}
\end{align*}
\end{lemma}
\begin{IEEEproof}
\begin{align*}
&\Prob{ \left. \max_{k \in S_m} \left| \phi_m^*h_k\right|^2 <  x\right| \left| S_m\right| \geq nq_n}\\
&= \sum_{s=\lceil n(\bar p-\delta) \rceil}^n 
\sum_{\substack{\mathcal{A}_s\subseteq \lb n\rb:\\ \left| \mathcal{A}_s\right|=s}}
\Prob{\left. S_m = \mathcal{A}_s\right|\left| S_m\right| \geq nq_n} \\
&\quad \cdot\Prob{ \left. \max_{k \in \mathcal{A}_s} \left| \phi_m^*h_k\right|^2 <  x\right| \left| \mathcal{A}_s\right| \geq nq_n, S_m=\mathcal{A}_s}\\
&\overset{\aaaa}{=} \sum_{s=\lceil nq_n \rceil}^n 
\Prob{ \left|S_m\right| = s\left|\left| S_m\right| \geq nq_n \right. } \\
&\qquad \cdot\Prob{ \left. \max_{k \in \mathcal{A}_s} \left| \phi_m^*h_k\right|^2 < x\right| \left| \mathcal{A}_s\right| =s}\\
&\overset{\bbbb}{\leq}  \Prob{ \left. \max_{k \in \mathcal{A}_s} \left| \phi_m^*h_k\right|^2 <  x\right| \left| \mathcal{A}_{s}\right| = nq_n}\\
&\overset{\cccc}{=}\lp  1-e^{ -\frac{x}{2}}\rp^{nq_n}
\end{align*}
where (a) follows by the fact that the probability on the right-hand side does not depend on $\mathcal{A}_s$ as long as $\left| \mathcal{A}_s\right|$ is fixed, owing to the fact that the user channel vectors $ h_k$ are i.i.d., and since $\Phi h_k \sim \mathcal{CN}(0,I)$;
(b) follows because the given probability is a monotonically decreasing function of $s$;
and (c) is because $\lbp \left| \phi_m^*h_k\right|^2 \rbp$ are i.i.d. $\chi^2(2)$ distributed;
\end{IEEEproof}

\begin{lemma}\label{lem:chi2}
Let $X_1,\dots,X_N$ be i.i.d. $\chi^2(2)$ distributed random variables. Then
\begin{align*}
\Prob{\max_{1\leq i \leq N} X_i < 2\log N-2\log\log N} \leq \frac{1}{N}
\end{align*}
\end{lemma}
\begin{IEEEproof}
\begin{align*}
&\Prob{\max_{1\leq i \leq N} X_i < 2\log N -\log\log N} \\
&= \lb\Prob{ X_1 < 2\log N -\log\log N}\rb^N\\
& =\lp 1-\exp\lbp -\log N + \log\log N \rbp\rp^N = \lp 1-\frac{\log N}{N}\rp^N\\
&=\exp\lbp N\log\lp 1 - \frac{\log N}{N}\rp\rbp \\
&= \exp\lbp N\lp -\frac{\log N}{N} -O\lp\frac{\log^2 N}{N^2}\rp\rp\rbp \leq \frac{1}{N}
\end{align*}
\end{IEEEproof}

\begin{lemma}\label{lem:chi2m}
Let $X_1,\dots,X_N$ be i.i.d. $\chi^2(2M)$ distributed random variables. Then for $N$ sufficiently large,
\begin{align*}
&\Prob{\max_{1\leq i \leq N} X_i > 2\log N + \lp 2M+2 \rp\log \log N} \\
&\qquad\qquad\qquad\qquad\qquad\qquad\qquad\qquad\qquad= O\lp \frac{1}{\log N}\rp
\end{align*}
\end{lemma}
\begin{IEEEproof}
Chernoff bound for a $\chi^2(2M)$ random variable $Z$ is given by 
\begin{align*}
\Prob{Z > x} \leq \lp \frac{x}{2M}e^{1-\frac{x}{2M}}\rp^M,
\end{align*}
for $x>2M$. Then, assuming $N$ is large enough,
\begin{align*}
&\Prob{\max_{1\leq i \leq N} X_i >  2\log N + \lp 2M+2 \rp\log \log N} \\
&= 1 - \Prob{\max_{1\leq i \leq N} X_i \leq  2\log N + \lp 2M+2 \rp\log \log N}\\
& = 1 - \lb\Prob{X_1 \leq  2\log N + \lp 2M+2 \rp\log \log N}\rb^N \\
& = 1 - \lb 1 -\Prob{X_1 >  2\log N + \lp 2M+2 \rp\log \log N}\rb^N \\
& \leq 1 - \lp 1 - \lp \frac{2\log N + \lp 2M+2 \rp\log \log N}{2M} \right.\right.\\
&\left.\left.\qquad\exp\lbp 1-\frac{2\log N + \lp 2M+2 \rp\log \log N}{2M}\rbp \rp^M\rp^N \\
& = 1 - \lp 1 -  \frac{\lp 2\log N + \lp 2M+2 \rp\log \log N\rp^M e^M}{\lp 2M\rp^{M}N\log^{M+1} N}\rp^N \\
&\doteq 1 - \exp\lbp - \frac{\lp 2\log N + \lp 2M+2 \rp\log \log N\rp^M}{\lp 2M/e\rp^{M}\log^{M+1} N} \rbp\\
&\overset{\aaaa}{\leq}  \lp\frac{e}{2M}\rp^M\frac{\lp 2\log N + \lp 2M+2 \rp\log \log N\rp^M}{\log^{M+1} N}\\
&=  \lp\frac{e}{2M}\rp^M\frac{O\lp \log^M N\rp}{\log^{M+1} N} = O\lp \frac{1}{\log N}\rp
\end{align*}
where (a) is by the inequality $1-x \leq e^{-x}$.
\end{IEEEproof}
\begin{lemma}\label{lem:Gn_epsilon}
If $N_n \to \infty$ and $\epsilon_n \to 0$ as $n\to\infty$, then for i.i.d. $\chi^2 (2)$ distributed $X_i,\dots,X_{N_n}$,
\begin{align*}
\lim_{n\to\infty} \Prob{\max_{1 \leq k \leq N_n} X_k > \epsilon_n} = 1.
\end{align*}
\end{lemma}
The proof for Lemma~\ref{lem:Gn_epsilon} is trivial and omitted here.